\documentclass[aps,preprint]{revtex4}%
\usepackage{amsfonts}
\usepackage{amsmath}
\usepackage{amssymb}
\usepackage{subfigure}
\usepackage{graphicx}%

\begin{document}
\title{Remarks on the weak cosmic censorship conjecture of RN-AdS
black holes with cloud of strings and quintessence under the scalar
field}
\author{Jing Liang$^{a,b}$}
\email{jingliang@stu.scu.edu.cn}

\author{Xiaobo Guo$^{c}$}
\email{guoxiaobo@czu.edu.cn}

\author{Deyou Chen$^{d}$}
\email{deyouchen@hotmail.com}

\author{Benrong Mu$^{a,b}$}
\email{benrongmu@cdutcm.edu.cn}

\affiliation{$^{a}$ Physics Teaching and Research section, College of Medical Technology,
Chengdu University of Traditional Chinese Medicine, Chengdu, 611137,
PR China}
\affiliation{$^{b}$Center for Theoretical Physics, College of Physics, Sichuan University, Chengdu, 610064, PR China}
\affiliation{$^{c}$Mechanical and Electrical Engineering School, Chizhou University, Chizhou, 247000, PR China}
\affiliation{$^{d}$School of Science, Xihua University, Chengdu, 610039, China}
\begin{abstract}
In this paper, we investigate the thermodynamics and weak cosmic censorship conjecture in a RN-AdS black hole with the cloud of strings and quintessence by the scattering of a scalar field. The variations of the thermodynamic variables are calculated in the normal and extended phase spaces. In the normal phase space, where the cosmological constant is considered as a constant, the first and second laws of thermodynamics are satisfied. In the extended phase space, where the cosmological constant and the parameters related to the cloud of strings and quintessence are treated as variables, the first law of thermodynamics is still satisfied and the second law of thermodynamics is indefinite. Moreover, we find that the weak cosmic censorship conjecture is valid for the extremal and near-extremal black holes in both phase spaces.
\end{abstract}
\keywords{}

\maketitle
\tableofcontents{}

\bigskip{}



\section{Introduction}

\label{sec:intro}

Recently, the existence of a gravitationally repulsive interaction at a global scale (cosmic dark energy) has been confirmed by high-precision observations. It is responsible for the accelerated expansion of the universe \cite{intro-Ade:2013sjv}. The latest results from the combination of cosmic microwave background (CMB) with type Ia supernova, large-scale structure (cosmic shear) and galaxy cluster abundance show that our universe is occupied by the mysterious dark energy with negative pressure (about $70\%$), contains cold dark matter with negligible pressure (about $25\%$) and the ordinary baryonic matter (about $5\%$) \cite{intro-Mukhanov:2005sc,intro-Ade:2015xua}. There are several types of dark energy models, such as cosmological constants \cite{intro-Cardenas:2002np}, k-essence \cite{intro-Yang:2009zzl} and quintom \cite{intro-Guo:2004fq,intro-Xia:2006rr} models. For one of the models, the equation of state parameter is in the range of $-1\leq\omega_q\leq-\frac{1}{3}$. This type of models is called quintessence dark energy or quintessence for short. In an astrophysical scenario, quintessence produces some gravitational effects when it surrounds black holes \cite{intro-Saleh:2011zz}. When the impact of quintessence in this scenario needs to be understood and further studied, it is necessary to solve the Einstein equation of this source. In recent years, Kiselev has considered Einstein's field equation surrounded by quintessence and obtained a new solution that depends on the quintessence state parameters \cite{intro-Kiselev:2002dx}.

According to string theory, nature can be represented by a set of extended objects (such as one-dimensional strings) rather than point particles. Therefore, it is of great importance to understand the gravitational effects caused by a set of strings, which can be achieved by solving Einstein's equations with a finite number of strings. Among the results obtained by Letelier \cite{intro-Letelier:1979ej}, most of the subsequent studies on this aspect focus on spherical symmetry. In this case, the Schwarzschild black hole is surrounded by a spherically symmetric cloud of strings. In other words, this situation corresponds to a modification of Schwarzschild's solution, which uses the cloud of strings as an additional source of the gravitational field. The existence of cloud of strings will produce a global origin effect that related to a solid deficit angle. The solid deficit angle depends on the existence of cloud \cite{intro-Barriola:1989hx}. In addition, the presence of clouds has a profound effect on the layered structure, which has a larger radius than the Schwarzschild radius. Therefore, the influence of global origin related to global deflection angle will have possible astrophysical consequences, which justifies the study of the cloud of strings in this case.

There have been some studies about black holes surrounded by the cloud of strings and quintessence \cite{ intro-Toledo:2020xnt, intro-Toledo:2018hav, intro-Ma:2019pya,intro-Chabab:2020ejk, intro-Sakti:2019iku, intro-Toledo:2019amt, intro-Toledo:2018pfy}. Some of them discussed about the thermodynamics of black holes. Since the works of Bekenstein \cite{intro-Bekenstein:1972tm,intro-Bekenstein:1973ur} and Hawking \cite{intro-Hawking:1974rv,intro-Hawking:1974sw,intro-Hawking:1976de}
in the 1970s, black hole thermodynamics has been an active and important research field.

An important feature of a black hole is that it has a horizon. There is a gravitational singularity in the center of the black hole, which hides in the event horizon. Moreover, according to Einstein’s theory of gravity, the existence of this singularity is inevitable \cite{intro-Hawking:1969sw}. Near the singularity, the laws of physics break down. In order to avoid this phenomenon, Penrose proposed the weak cosmic censorship conjecture (WCCC) in 1969 \cite{intro-Penrose:1964wq,intro-Penrose:1969pc}. The weak cosmic censorship conjecture asserts that the singularity is hidden behind event horizon and cannot be seen by distant observers. Although this conjecture seems reasonable, there is a lack of concrete evidence and people can only test its validity. The Gedanken experiment is an effective method to test the validity of the weak cosmic censorship conjecture \cite{intro-Sorce:1974dst}. In this experiment, a test particle with sufficient large energy, charge and angular momentum is thrown into the black hole to observe whether the horizon of the black hole is destroyed after the particle is absorbed. If the horizon is destroyed, the singularity is not surrounded by the horizon and becomes a naked singularity. If the horizon isn't destroyed, the singularity is surrounded by the horizon. People have used this experiment to study the validity of the weak cosmic censorship conjecture in various black holes since it was proposed \cite{intro-Ying:2020bch,intro-Yang:2020czk,intro-Mu:2019bim,intro-Hu:2019lcy,intro-He:2019fti,intro-Wang:2019dzl,intro-Liu:2020cji,intro-Hu:2020lkg,intro-Wang:2020osg,intro-Wang:2019jzz,intro-Chen:2020zps,intro-Hu:2020ccj,intro-Zeng:2019baw,intro-Zeng:2019hux,intro-Zeng:2019aao,intro-Han:2019lfs,intro-Han:2019kjr,intro-Zeng:2019jrh}. In addition, the validity of the weak cosmic censorship conjecture can be studied through the Gedanken experiment by using test fields instead of test particles. This experiment was first proposed by Semiz \cite{intro-Semiz:2005gs}. People also studied the validity of the weak cosmic censorship conjecture in various space-time by this experiment \cite{intro-Hong:2020zcf,intro-Yang:2020iat,intro-Bai:2020ieh,intro-Hong:2019yiz,intro-Chen:2019nsr,intro-Chen:2018yah,intro-Gwak:2018akg,intro-Toth:2011ab,intro-Goncalves:2020ccm,intro-Jiang:2020btc}.

In this paper, we investigate the thermodynamics and weak cosmic censorship conjecture in a RN-AdS black hole with cloud of strings and quintessence by the scattering of a complex scalar field. Due to the existence of the cloud of strings and quintessence, the parameters related to the cloud of strings and quintessence are considered as extensive thermodynamic parameters and taken into account in the calculation\cite{intro-Liu:2017baz}.
The rest of this paper is organized as follows. In Section \ref{sec:M}, we review the thermodynamics of the black
hole. In Section \ref{sec:research}, we get the changes
in the internal energy and charge of the black hole during the time interval $dt$. In Section
\ref{sec:nor} and \ref{sec:extend}, the first and the
second law of thermodynamics and the weak cosmic censorship conjecture are discussed
in the normal and extended phase space, respectively. Our results are summarized
in Section \ref{sec:con}.

\section{Thermodynamics in the RN-AdS black hole surrounded by quintessence and cloud of strings}

\label{sec:M}

Recently, it was obtained the solution corresponding to a black hole with quintessence and cloud of strings. Assuming that the cloud of strings and quintessence do not interact, the energy-momentum tensor of the two sources can be seen as a linear superposition which is given by \cite{M-Toledo:2019amt}
\begin{equation}
T_{t}^{t}=T_{r}^{r}=\frac{a}{r^{2}}+\rho_{q},T_{\theta}^{\theta}=T_{\phi}^{\phi}=-\frac{1}{2}\rho_{q}\left(3\omega_{q}+1\right).
\end{equation}
In the above equation, the pressure and density of quintessence are related by the
equation of state $p_{q}=\rho_{q}\omega_{q}$, where $\omega_{q}$
is the quintessential state parameter. Considering the static spherically symmetric line element, the line element associated with the charged RN-AdS black hole surrounded by cloud of strings and quintessence is written as \cite{intro-Chabab:2020ejk,intro-Ma:2019pya,M-Singh:2020tkf}
\begin{equation}
ds^{2}=f(r)dt^{2}-\frac{1}{f(r)}dr^{2}-r^{2}\left(d\theta^{2}+sin^{2}\theta d\phi^{2}\right),
\end{equation}
with the electromagnetic potential
\begin{equation}
A_{\mu}(r)=(-\frac{Q}{r},0,0,0),
\end{equation}
where
\begin{equation}
f(r)=1-a-\frac{2M}{r}+\frac{Q^{2}}{r^{2}}-\frac{\alpha}{r^{3\omega_{q}+1}}-\frac{\text{\ensuremath{\Lambda}}r^{2}}{3},
\label{eqn:fr}
\end{equation}
$M$ and $Q$ are the mass and electric charge of the black hole, respectively. $\Lambda$ is the cosmological constant and $a$ is an integration constant related to the presence of the cloud of strings. $\omega_{q}$ is the state parameter which has that $-1\leq\omega_q\leq-\frac{1}{3}$ for the quintessence and $\alpha$ is the
normalization factor related to the density of quintessence as
\begin{equation}
\rho_{q}=-\frac{\alpha}{2}\frac{3\omega_{q}}{r^{3(\omega_{q}+1)}}.
\end{equation}

\begin{figure}[htb]
\begin{center}
\subfigure[{$\alpha=0.01$,$\omega_{q}=-2/3$.}]{
\includegraphics[width=0.45\textwidth]{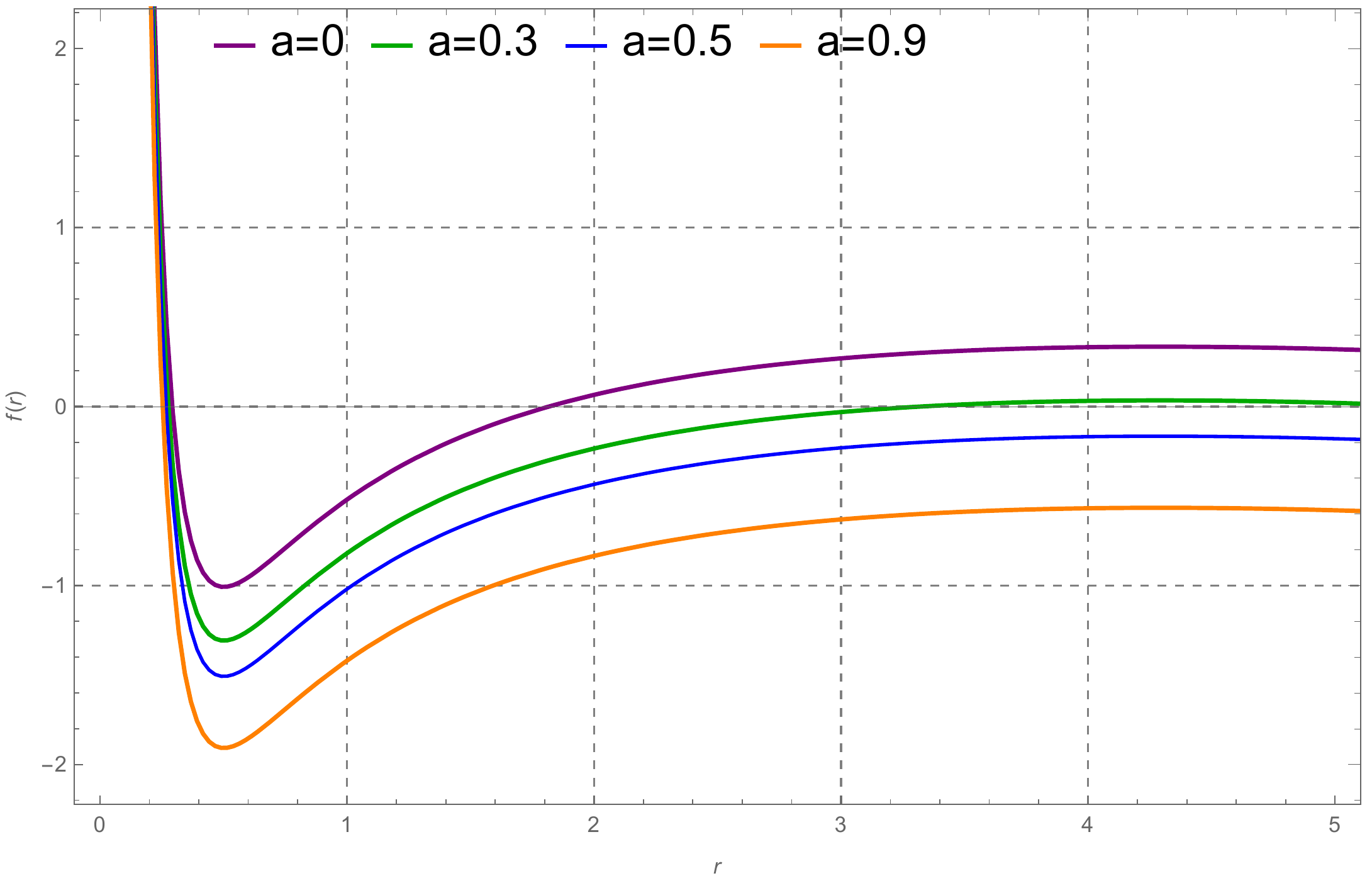}\label{fig:f1}}
\subfigure[{$\alpha=0.1$,$\omega_{q}=-2/3$.}]{
\includegraphics[width=0.45\textwidth]{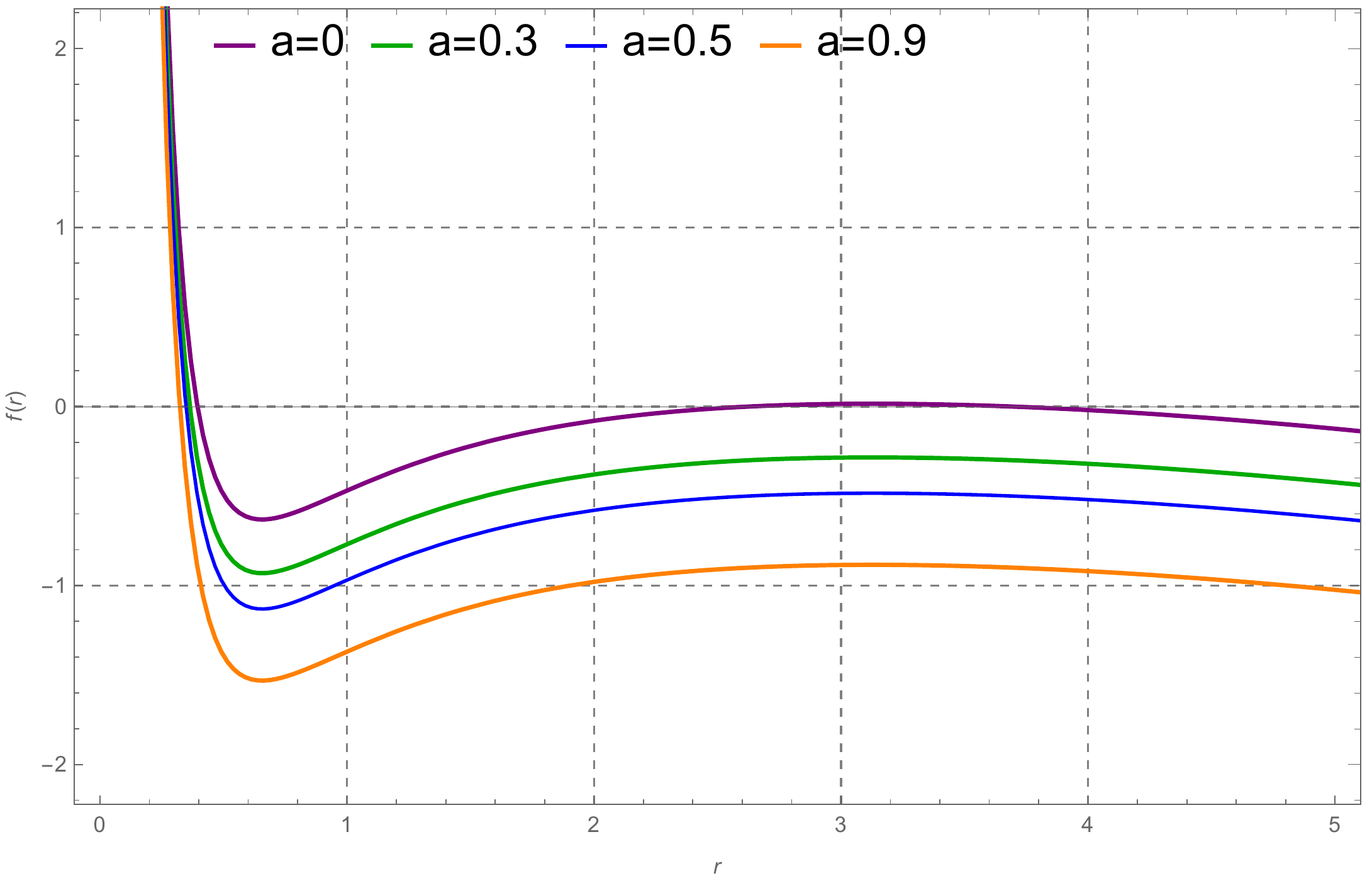}\label{fig:f2}}
\subfigure[{$\alpha=0.01$,$\omega_{q}=-1/6$.}]{
\includegraphics[width=0.45\textwidth]{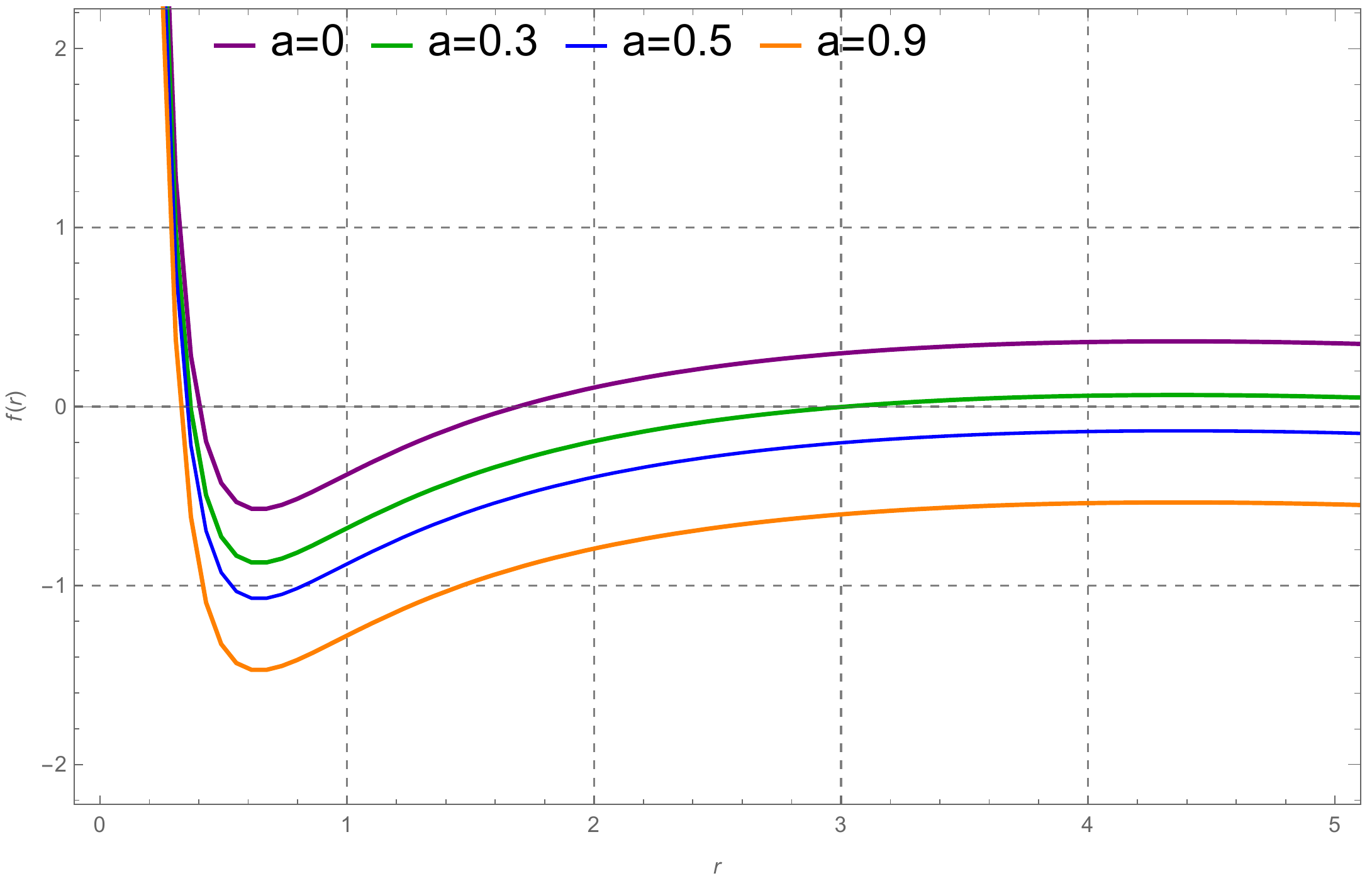}\label{fig:f3}}
\subfigure[{$\alpha=0.1$,$\omega_{q}=-1/6$.}]{
\includegraphics[width=0.45\textwidth]{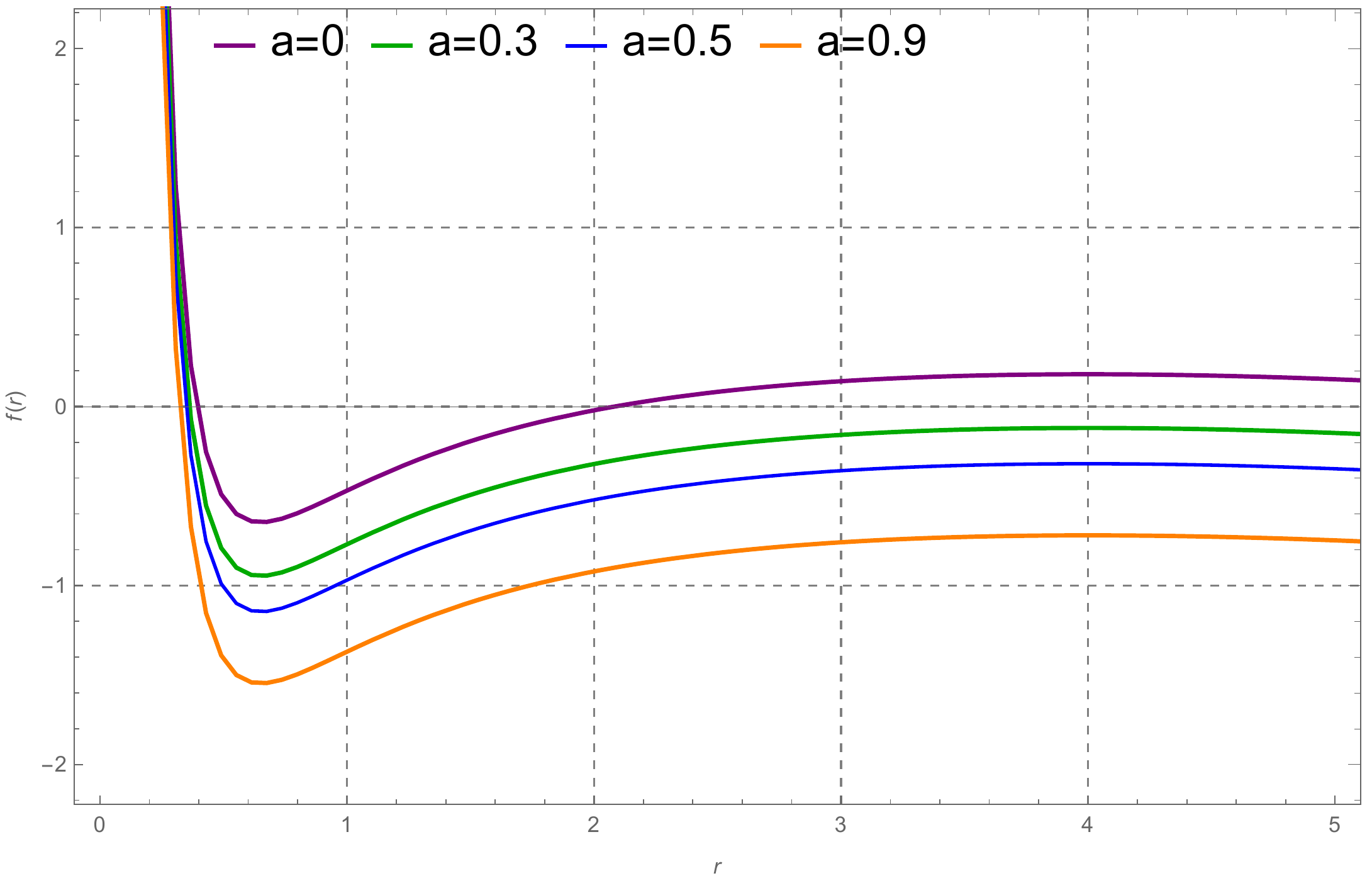}\label{fig:f4}}
\end{center}
\caption{The function $f(r)$ for different values of $a$, $\alpha$ and $\omega_{q}$. We choose $M=1$ and $Q=0.8$.}%
\label{fig:f}
\end{figure}

In Fig. \ref{fig:f}, the graphs of the function $f(r)$ are shown for different
values of the parameters $a$, $\alpha$ and $\omega_{q}$. In Fig. \ref{fig:f}, we can intuitively observe the effect due to the cloud of strings and quintessence. The value of $f(r)$ increases when $a$ or $\omega_{q}$ increases. Besides, when the value of $\alpha$ increases, $f(r)$ decreases.

When the black hole is non-extremal, equation $f(r)=0$ has two positive real roots $r_-$ and $r_+$, which represent the radius of the Cauchy horizon and the event horizon, respectively. When the black
hole is extremal, $f(r)=0$ only has a single root $r_+$.
The Hawking temperature of the black hole is
\begin{equation}
T=\frac{f^{\prime}(r_{+})}{4\pi}=\frac{1}{4\pi}(\frac{2M}{r_{+}^{2}}-\frac{2Q^{2}}{r_{+}^{3}}+\frac{(3\omega_{q}+1)\alpha}{r_{+}^{3\omega_{q}+2}}+\frac{2r_{+}}{l^{2}}),
\end{equation}
The electric potential and the entropy take on the form as
\begin{equation}
\varphi=\frac{Q}{r_{+}},
\label{eqn:potential}
\end{equation}
\begin{equation}
S=\pi r_{+}^{2}.
\label{eqn:entropy}
\end{equation}
The black hole can be discussed in the normal phase space and the extended phase space as a stable thermodynamic system. In the normal space, the cosmological constant is treated as a constant. Then the first law
of thermodynamics is written as
\begin{equation}
dM=TdS+\varphi dQ.
\end{equation}
In the extended phase space, the cosmological constant is treated as a function of the pressure. The relation between the cosmological constant and the pressure is given by \cite{M-Dolan:2011xt,M-Kubiznak:2012wp,M-Cvetic:2010jb,M-Caceres:2015vsa,M-Hendi:2012um,M-Pedraza:2018eey}
\begin{equation}
P=-\frac{\Lambda}{8\pi}=\frac{3}{8\text{\ensuremath{\pi}}l^{2}}.
\label{eqn:P V}
\end{equation}
Then the first law of thermodynamics takes on the form as
\begin{equation}
dM=TdS+VdP+\text{\ensuremath{\varphi}}dQ+\gamma\text{\ensuremath{d\alpha}}+\varkappa da,
\end{equation}
where
\begin{equation}
\gamma=-\frac{1}{2r_{+}^{3\omega_{q}}},\varkappa=-\frac{r_{+}}{2}.
\end{equation}
The mass of the black hole $M$ is defined as its enthalpy. Hence, the relationship between enthalpy, internal energy and pressure is \cite{M-Cvetic:2010jb,M-Kastor:2009wy}
\begin{equation}
M=U+PV.
\label{eqn:M}
\end{equation}

\section{Energy and charge's variation of the RN-AdS black hole surrounded by quintessence and cloud of strings}

\label{sec:research}
When the space-time dimension is assumed to be four, the action of the charged scalar field in the fixed gravitational and electromagnetic fields is \cite{re-Gwak:2019asi}
\begin{equation}
S=-\frac{1}{2}\int\sqrt{-g}\left(\psi^{\ast}\psi D^{\mu}D_{\mu}-m^{2}\psi^{\ast}\psi\right)d^{4}x.
\label{eqn:S}
\end{equation}
Due to the scalar field with charge $q$, we consider that $D^{\mu}=\partial^{\mu}-iqA^{\mu}$ and $D_{\mu}=\partial_{\mu}+iqA_{\mu}$. In the above equation, $m$ is the mass, $q$ is the charge, $A_{\mu}$ is the electromagnetic potential and $\psi$ is the the wave function whose conjugate is $\psi^{\ast}$. The field equation get from the
action satisfies
\begin{equation}
\left(\nabla^{\mu}-iqA^{\mu}\right)(\nabla_{\mu}+iqA_{\mu})\psi-m^{2}\psi=0.
\end{equation}
To solve this equation, it is necessary to use the separation of variables
\begin{equation}
\psi=e^{-i \omega t}R(r)\varPhi(\theta,\phi),
\end{equation}
where $\omega$ is the energy of the particle and $\varPhi(\theta,\phi)$
is the scalar spherical harmonics. At the outer horizon, the
radial solution of the scalar field is
\begin{equation}
R(r)=e^{\pm i\left(\omega-\frac{qQ}{r}\right)r_{\ast}}.
\label{eqn:R}
\end{equation}
In the above equation, $dr_{\ast}=\frac{1}{f}dr$ and $+/-$ corresponds to the
solution of the ingoing/outgoing radial wave. Since this article discusses the thermodynamics and validity of the weak cosmic censorship conjecture through the scattering of ingoing waves in the event horizon, we focus on the ingoing wave
function.
From Eq. $\left(\ref{eqn:S}\right)$, the energy-momentum tensor is written as
\begin{equation}
T_{\nu}^{\mu}=\frac{1}{2}[\left(\partial^{\mu}-iqA^{\mu}\right)\psi^{\ast}\partial_{\nu}\psi+\left(\partial^{\mu}+iqA^{\mu}\right)\psi\partial_{\nu}\psi^{\ast}]+\delta_{\nu}^{\mu}\mathcal{L}.
\end{equation}
The energy flux is the component $T_{\nu}^{\mu}$ integrated by a solid angle on an $S^2$ sphere at the outer horizon. The energy flux is given by
\begin{equation}
\frac{dE}{dt}=\int T_{t}^{r}\sqrt{-g}d\theta d\phi=\omega(\omega-q\varphi)r_{+}^{2}.
\end{equation}
The charge flux can get from the energy flux \cite{re-Bekenstein:1973mi}. Therefore, the charge flux is
\begin{equation}
\frac{dQ}{dt}=-\int j^{r}\sqrt{-g}d\theta d\varphi=q\left(\omega-q\varphi\right)r_{+}^{2}.
\end{equation}
The fluxes in the above formulas will change the corresponding
characteristics of the black hole in the infinitesimal time interval $dt$. The charge flux corresponds to the change in that of the black hole. Moreover, the energy flux corresponds to the internal energy of the black hole. Then the changes in internal energy and charge are
\begin{equation}
dU=dE=\omega(\omega-q\varphi)r_{+}^{2}dt,dQ=q(\omega-q\text{\ensuremath{\varphi}})r_{+}^{2}dt.
\label{eqn:dUdQ}
\end{equation}

\section{Thermodynamics and the weak cosmic censorship conjecture in the normal space}

\label{sec:nor}

\subsection{Thermodynamics in the normal phase space}

\label{subsec:nor T}

During the scattering of the scalar field, the mass $M$, the charge $Q$ and other thermodynamic variables of the black bole change due to the conservation law. Assuming that the black hole's initial state is expressed by $(M,Q,r_{+})$
and final state is expressed by $(M+dM,Q+dQ,r_{+}+dr_{+})$. The initial state $(M,Q,r_{+})$ and the final state $(M+dM,Q+dQ,r_{+}+dr_{+})$ satisfy
\begin{equation}
f\left(M,Q,r_{+}\right)=f\left(M+dM,Q+dQ,r_{+}+dr_{+}\right)=0.
\label{eqn:nor T f}
\end{equation}
The functions $f\left(M,Q,r_{+}\right)$ and $f\left(M+dM,Q+dQ,r_{+}+dr_{+}\right)$ satisfy
the following relation
\begin{equation}
\begin{aligned}
&f\left(M+dM,Q+dQ,r_{+}+dr_{+}\right)=f\left(M,Q,r_{+}\right)\\
&+\frac{\partial f}{\partial M}|_{r=r_{+}}dM+\frac{\partial f}{\partial Q}|_{r=r_{+}}dQ+\frac{\partial f}{\partial r}|_{r=r_{+}}dr_{+},\\
\end{aligned}
\label{eqn:nor T f1 eqn}
\end{equation}
where
\begin{equation}
\frac{\partial f}{\partial M}|_{r=r_{+}}=-\frac{2}{r_{+}},\frac{\partial f}{\partial Q}|_{r=r_{+}}=\frac{2Q}{r_{+}^{2}},\frac{\partial f}{\partial r}|_{r=r_{+}}=4\pi T.
\label{eqn:nor T df}
\end{equation}
Substituting Eqs. $\left(\ref{eqn:potential}\right)$, $\left(\ref{eqn:entropy}\right)$, $\left(\ref{eqn:nor T f}\right)$ and
$\left(\ref{eqn:nor T df}\right)$ into Eq. $\left(\ref{eqn:nor T f1 eqn}\right)$, we obtain
\begin{equation}
dM=TdS+\varphi dQ.
\label{eqn:nor dM}
\end{equation}
This is the first law of thermodynamics of the RN-AdS black hole with cloud of strings and quintessence in the normal phase space.
From Eqs. $\left(\ref{eqn:nor T f}\right)$, $\left(\ref{eqn:nor T f1 eqn}\right)$ and $\left(\ref{eqn:nor T df}\right)$, the variation of the radius in horizon takes on the form as
\begin{equation}
dr_{+}=\frac{r_{+}}{2\text{\ensuremath{\pi}}T}(\omega-q\varphi)^{2}dt.
\end{equation}
Hence, the variation of entropy is
\begin{equation}
dS=2\text{\ensuremath{\pi}}r_{+}dr_{+}=\frac{r_{+}^{2}(\omega-q\varphi)^{2}}{T}dt>0.
\end{equation}
Therefore, the entropy of the black hole increases and the second law of thermodynamics is satisfied in the normal phase space.

\subsection{Weak cosmic censorship conjecture in the normal phase space}

\label{subsec:nor W}

\begin{figure}[htb]
\begin{center}
\subfigure[{$f\left( r\right)$ in naked singularities.}\label{fig:WCCC1}]{
\includegraphics[width=0.3\textwidth]{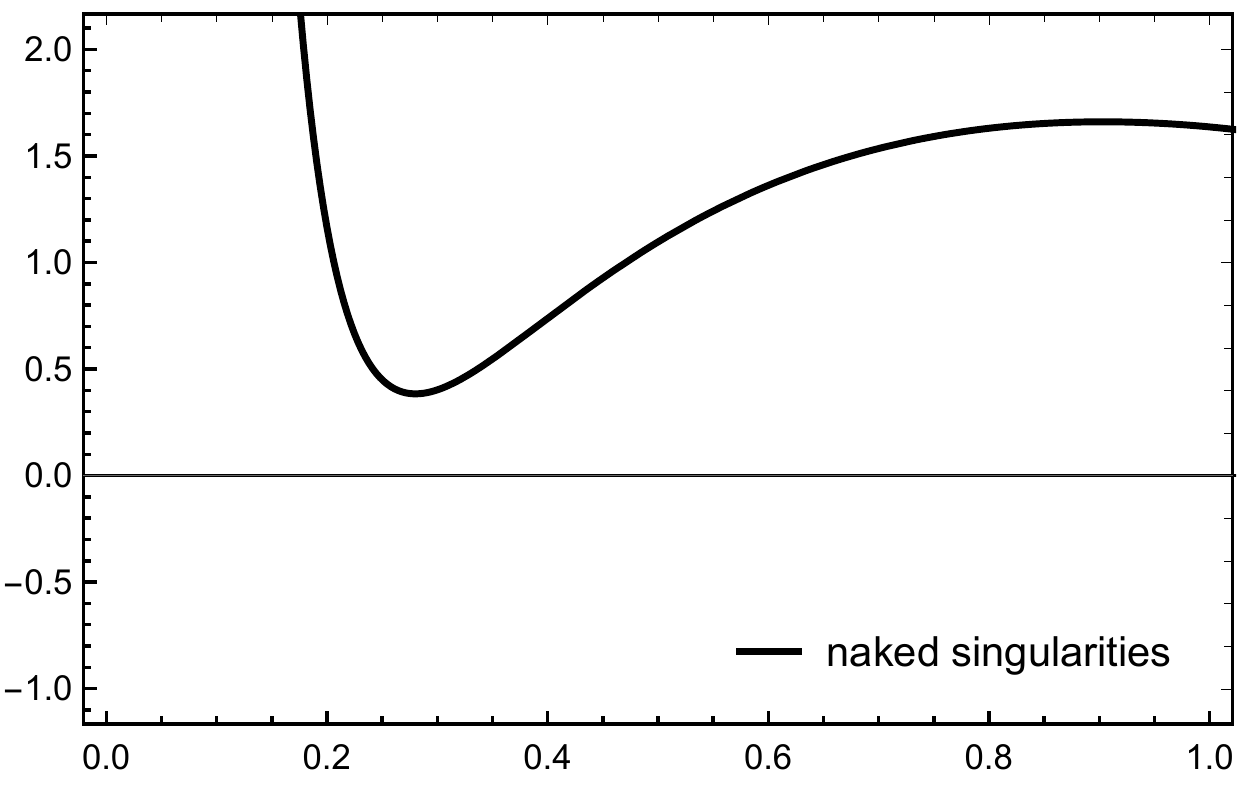}}
\subfigure[{$f\left( r\right)$ in non-extremal black holes.}\label{fig:WCCC2}]{
\includegraphics[width=0.3\textwidth]{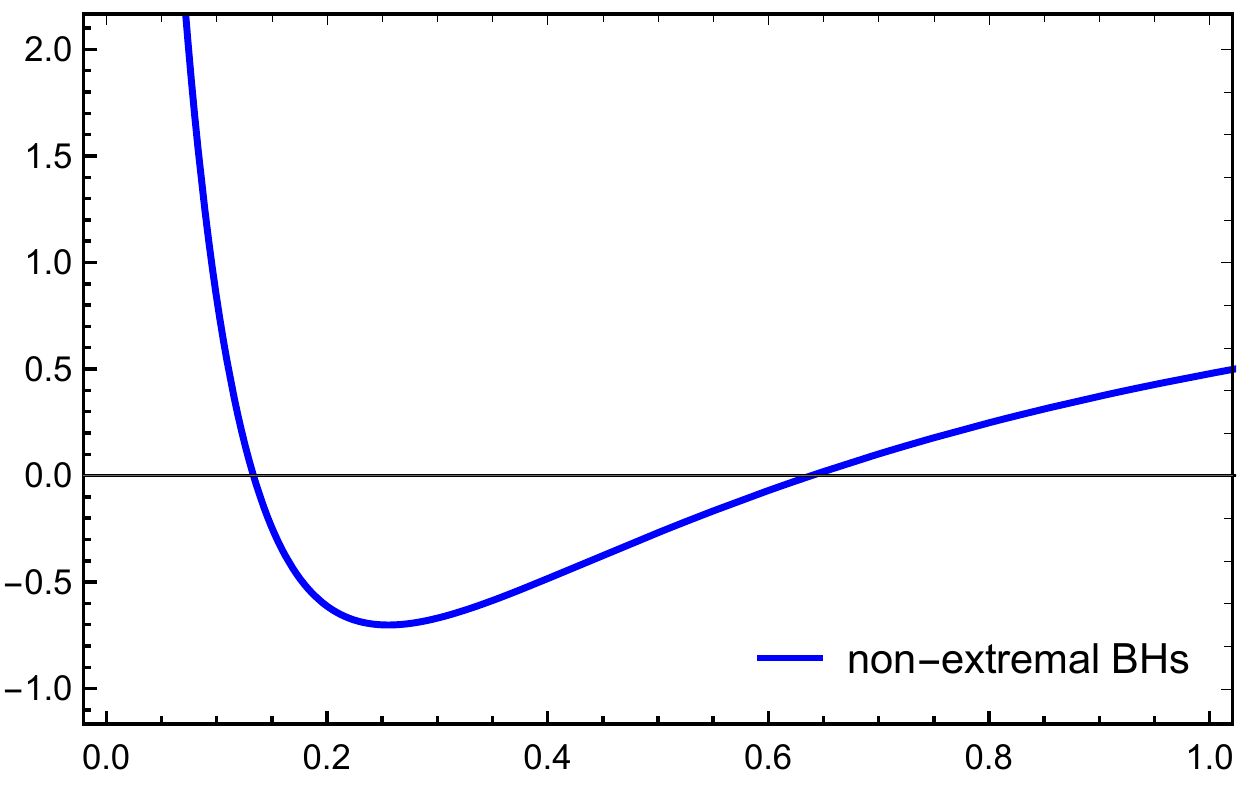}}
\subfigure[{$f\left( r\right)$ in extremal black holes.}\label{fig:WCCC3}]{
\includegraphics[width=0.3\textwidth]{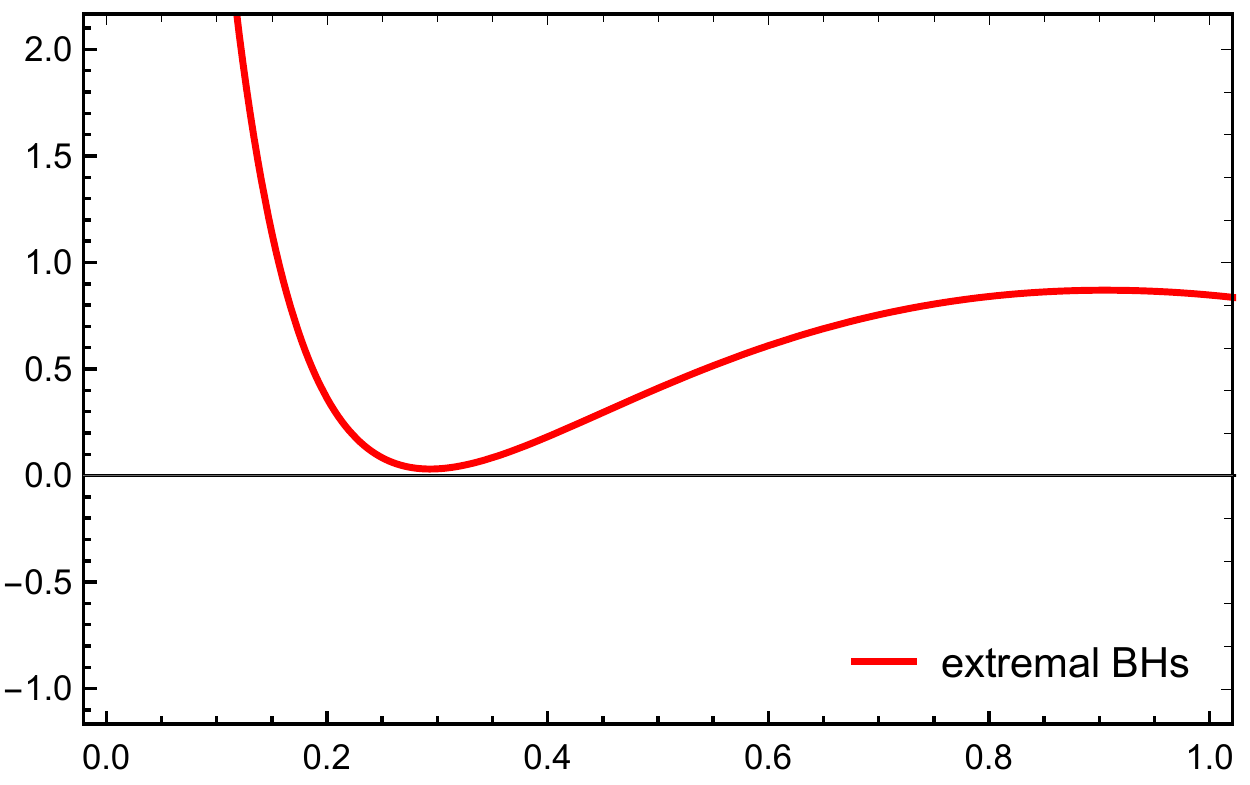}}
\end{center}
\caption{Graphs of $f\left( r\right)$ for given states of the RN-AdS black holes with cloud of strings and quintessence.}%
\label{fig:WCCC}
\end{figure}

In this section, the stability of the event horizon under the scattering
of a scalar field in the normal phase space is investigated. In the initial
state, the minimum value of the function $f(r)$ is negative or zero, $f(r)=0$ has real roots and event horizon exists. When
the flux of the scalar field enters the black hole, the mass and charge
of the black hole change during the time interval $dt$. Moreover, the minimum value of $f(r)$ also changes. If the minimum
value is positive, as shown in Fig. \ref{fig:WCCC1}, no solution represents the horizon in the final state and the singularity becomes a naked singularity. In this case, the black
hole is overcharged and the weak cosmic censorship conjecture is violated. On the contrary, if the minimum value is negative or equal to zero in the final state, as
shown in Fig. \ref{fig:WCCC2} and Fig. \ref{fig:WCCC3}, at least one solution
representing the horizon in the final state and the singularity is covered by the horizon. The weak cosmic censorship conjecture is valid. Therefore, checking the sign of the minimum value in the final state is an effective way to test the validity of weak cosmic censorship conjecture after scalar field scattering.

The sign of the minimum value of $f(r)$ in the
final state can be obtained in term of the initial state \cite{re-Gwak:2019asi}. In the initial state, $f(r)$ has a minimum value at $r=r_{\min}$. Moreover, the minimum value of $f(r)$ is not
greater than zero
\begin{equation}
f(r)_{min}=f(r,M,Q,)|_{r=r_{min}}=\delta\leq0.
\end{equation}
When the flux of the scalar field enters the black hole, the minimum point $r_{min}$ moves to $r_{min}+dr_{min}$.
Then for the final state, the minimum value of $f(r)$ is written as
\begin{equation}
f\left(M+dM,Q+dQ,r_{min}+dr_{min}\right)=\delta+\frac{\partial f}{\partial M}|_{r=r_{min}}dM+\frac{\partial f}{\partial Q}|_{r=r_{min}}dQ+\frac{\partial f}{\partial r}|_{r=r_{min}}dr_{min},
\label{eqn:nor W f min}
\end{equation}
where
\begin{equation}
\frac{\partial f}{\partial M}|_{r=r_{min}}=-\frac{2}{r_{min}},\frac{\partial f}{\partial Q}|_{r=r_{min}}=\frac{2Q}{r_{min}^{2}},\frac{\partial f}{\partial r}|_{r=r_{min}}=0.
\label{eqn:nor W df}
\end{equation}
Substituting Eqs. $\left(\ref{eqn:nor dM}\right)$ and $\left(\ref{eqn:nor W df}\right)$ into Eq. $\left(\ref{eqn:nor W f min}\right)$, we obtain
\begin{equation}
f\left(M+dM,Q+dQ,r_{min}+dr_{min}\right)=\delta-\frac{2}{r_{min}}(\omega-\frac{qQ}{r_{+}})(\omega-\frac{qQ}{r_{min}})r_{+}^{2}dt.
\label{eqn:nor W f}
\end{equation}

When it's an extremal black hole, $\delta=0$ and $r_{+}=r_{min}$, then Eq. $\left(\ref{eqn:nor W f}\right)$ becomes
\begin{equation}
f\left(M+dM,Q+dQ,r_{min}+dr_{min}\right)=-2(\omega-q\text{\ensuremath{\varphi}})^{2}r_{+}dt.
\end{equation}
In this case, the black hole can not be overcharged in the final state and the weak cosmic censorship
conjecture is valid. If $\omega=q\varphi$, the extremal black hole still extremal black hole. If $\omega\neq q\varphi$,
the extremal black hole will change to non-extremal black hole.

When it is a near-extremal black hole, $\delta$ is a small negative quantity and $r_{+}$ is numerically greater than $r_{min}$. The function $f(r_{\min}+dr_{\min},M+dM,Q+dQ)$ has the maximum value when $\omega=\frac{1}{2}qQ(\frac{1}{r_{+}}+\frac{1}{r_{min}})$. The maximum value is
\begin{equation}
f(M+dM,Q+dQ,r_{min}+dr_{min})=\delta-\frac{qQ^{2}\left(r_{+}-r_{min}\right)^{2}}{2r_{min}^{3}}dt<0.
\end{equation}
Hence, the minimum value of $f(r)$ is always less than zero. The event horizon exists in the final stage and the black hole is not overcharged after the scalar field
scattering. Therefore, the weak cosmic censorship conjecture
is valid in a near-extremal RN-AdS black hole surrounded by cloud
of strings and quintessence in the normal space.

\section{Thermodynamic and weak cosmic censorship conjecture in the extended phase space}

\label{sec:extend}

\subsection{Thermodynamics in the extended phase space}

\label{subsec:ext T}

In the extended phase space, the cosmological
constant is treated as a thermodynamic
variable. In order to consider the effect of cloud of strings and quintessence, the parameters related to the cloud of strings and quintessence are considered as extensive thermodynamic parameters. The black hole's initial state is expressed by $(M,Q,l,a,\alpha,r_{+})$
and final state is expressed by $(M+dM,Q+dQ,l+dl,a+da,\alpha+d\alpha,r_{+}+dr_{+})$.
The initial state $(M,Q,l,a,\alpha,r_{+})$ satisfies
\begin{equation}
f(M,Q,l,a,\alpha,r_{+})=0.
\label{eqn:ext T f}
\end{equation}
Assuming that the black hole's final state is still a
black hole, it satisfies
\begin{equation}
f\left(M+dM,Q+dQ,l+dl,a+da,\alpha+d\alpha,r_{+}+dr_{+}\right)=0.
\label{eqn:ext T f1}
\end{equation}
The functions $f(M+dM,Q+dQ,l+dl,a+da,\alpha+d\alpha,r_{+}+dr_{+})$ and $f(M,Q,l,a,\alpha,r_{+})$ satisfy
\begin{equation}
\begin{aligned}
& f\left(M+dM,Q+dQ,l+dl,a+da,\alpha+d\alpha,r_{+}+dr_{+}\right)=f\left(M,Q,l,a,\alpha,r_{+}\right)\\
& +\frac{\partial f}{\partial M}|_{r=r_{+}}dM+\frac{\partial f}{\partial Q}|_{r=r_{+}}dQ+\frac{\partial f}{\partial l}|_{r=r_{+}}dl+\frac{\partial f}{\partial a}|_{r=r_{+}}da\\
&+\frac{\partial f}{\partial\alpha}|_{r=r_{+}}d\text{\ensuremath{\alpha}}+\frac{\partial f}{\partial r}|_{r=r_{+}}dr_{+},
\end{aligned}
\label{eqn:ext T f1eqn}
\end{equation}
where
\begin{equation}
\begin{aligned}
& \frac{\partial f}{\partial M}|_{r=r_{+}}=-\frac{2}{r_{+}},\frac{\partial f}{\partial Q}|_{r=r_{+}}=\frac{2Q}{r_{+}^{2}},\frac{\partial f}{\partial l}|_{r=r_{+}}=-\frac{2r_{+}^{2}}{l^{3}},\\
 & \frac{\partial f}{\partial\alpha}|_{r=r_{+}}=-\frac{1}{r_{+}^{3\omega_{q}+1}},\frac{\partial f}{\partial a}|_{r=r_{+}}=-1,\frac{\partial f}{\partial r}|_{r=r_{+}}=4\text{\ensuremath{\pi}}T.\\
\end{aligned}
\label{eqn:ext T df}
\end{equation}
Substituting Eqs. $\left(\ref{eqn:potential}\right)$, $\left(\ref{eqn:entropy}\right)$, $\left(\ref{eqn:P V}\right)$, $\left(\ref{eqn:ext T f}\right)$, $\left(\ref{eqn:ext T f1}\right)$
and $\left(\ref{eqn:ext T df}\right)$ into Eq. $\left(\ref{eqn:ext T f1eqn}\right)$,
we obtain
\begin{equation}
dM=TdS+VdP+\text{\ensuremath{\varphi}}dQ+\mathcal{\gamma}d\alpha+\varkappa da,
\label{eqn:ext T dM}
\end{equation}
where $\gamma$ and $\varkappa$ are conjugated
to the parameters $\text{\ensuremath{\alpha}}$ and $a$ which satisfy
\begin{equation}
\gamma=-\frac{1}{2r_{+}^{3\omega_{q}}},\varkappa=-\frac{r_{+}}{2}.
\end{equation}
When the parameters $\text{\ensuremath{\alpha}}$ and $a$ are treated as constants, the first law of thermodynamics
is written as
\begin{equation}
dM=TdS+\text{\ensuremath{\varphi}}dQ+VdP.
\end{equation}
In the following calculations, $a$ and $\alpha$ are still considered as variables to discuss the effect of the quintessence and the cloud of strings. Substituting Eq. $\left(\ref{eqn:P V}\right)$ into Eq. $\left(\ref{eqn:M}\right)$, we obtain
\begin{equation}
dM=d(PV+U)=\omega(\omega-q\text{\ensuremath{\varphi}})r_{+}^{2}dt+\frac{3r_{+}^{2}}{2l^{2}}dr_{+}-\frac{r_{+}^{3}}{l^{3}}dl.
\label{eqn:ext dM1}
\end{equation}
From Eqs. $\left(\ref{eqn:ext T dM}\right)$ and $\left(\ref{eqn:ext dM1}\right)$, the variation of radius takes on the form as
\begin{equation}
dr_{+}=\frac{2r_{+}l^{2}(\omega-q\text{\ensuremath{\varphi}})^{2}}{4\text{\ensuremath{\pi}}Tl^{2}-3r_{+}}dt+\frac{l^{2}r_{+}^{-3\omega_{q}-1}}{4\text{\ensuremath{\pi}}Tl^{2}-3r_{+}}d\alpha+\frac{l^{2}}{4\text{\ensuremath{\pi}}Tl^{2}-3r_{+}}da.
\end{equation}
Then the variation of entropy is
\begin{equation}
dS=2\text{\ensuremath{\pi}}r_{+}dr_{+}=\frac{4\text{\ensuremath{\pi}}r_{+}^{2}l^{2}(\omega-q\text{\ensuremath{\varphi}})^{2}}{4\text{\ensuremath{\pi}}l^{2}T-3r_{+}}dt+\frac{2\text{\ensuremath{\pi}}l^{2}r_{+}^{-3\omega_{q}}}{4\text{\ensuremath{\pi}}l^{2}T-3r_{+}}d\alpha+\frac{2\text{\ensuremath{\pi}}r_{+}l^{2}}{4\text{\ensuremath{\pi}}l^{2}T-3r_{+}}da.
\label{eqn:ext dS}
\end{equation}
When it is the extremal black hole, the temperature is zero. Then Eq. $\left(\ref{eqn:ext dS}\right)$ is written as
\begin{equation}
dS=-\frac{4}{3}l^{2}\text{\ensuremath{\pi}}r_{+}(\omega-q\text{\ensuremath{\varphi}})^{2}dt-\frac{2\pi}{3}l^{2}r_{+}^{-3\omega_{q}-1}d\alpha-\frac{2\text{\ensuremath{\pi}}}{3}l^{2}da.
\label{eqn:ext extremal dS}
\end{equation}
If $d\alpha>0$ and $da>0$, the entropy of the black hole decreases. If $d\alpha<0$ and $da<0$, the entropy of the black hole may increase. Therefore, the second law of thermodynamics
is indefinite for the extremal black hole in the extended phase space.

Then we focus on the near-extremal black hole. In order to intuitively understand the changes in entropy, we do numerically research on the change of entropy. We set $M=0.5$ and $l=1$. For the case $\omega_{q}=-2/3$, $a=0.1$ and $\alpha=0.1$, the extremal charge is $Q_e=0.487372428447$. When the charge is less than the extreme value, we use different charge values to produce the change in entropy. In Table \ref{tab:dS1}, we set $da=0.5$, $d\alpha=0.1$ and $dt=0.0001$. As can be seen from Table \ref{tab:dS1}, when the charge of the black hole decreases, the radius of event horizon increases, so does the value of $dS$. What is clear is that there are two regions, $dS<0$ and $dS>0$. This implies the existence of a phase change point, which divides $dS$ into positive and negative values.
\begin{table}[htb]
\begin{centering}
\begin{tabular}{ccccc}
\hline
\hline
$Q $               &$r_+ $               & $dS $       &$da$      &$d\alpha$  \\
\hline
0.487372428447    & 0.420154956933     & $-1.13539$      &          &              \\
0.487372          & 0.420632           & $-1.14446$      &          &              \\
0.48              & 0.480899           & $-4.13900$     &          &              \\
0.46              & 0.533323           & $20.0200$      &          &              \\
0.44              & 0.565356           & $6.16798$      &0.5       &0.1        \\
0.40              & 0.609271           & $3.76922$      &          &              \\
0.30              & 0.673541           & $2.85276$      &          &             \\
0.20              & 0.708703           & $2.65475$      &          &              \\
0.10              & 0.727240           & $2.58710$       &          &              \\
\hline
\hline
\end{tabular}
\par\end{centering}
\caption{{\footnotesize{}{}{}{}The relation between $dS$, $Q$ and $r_+$.}}
\label{tab:dS1}
\end{table}

\begin{figure}[htb]
\centering
\includegraphics[scale=0.8]{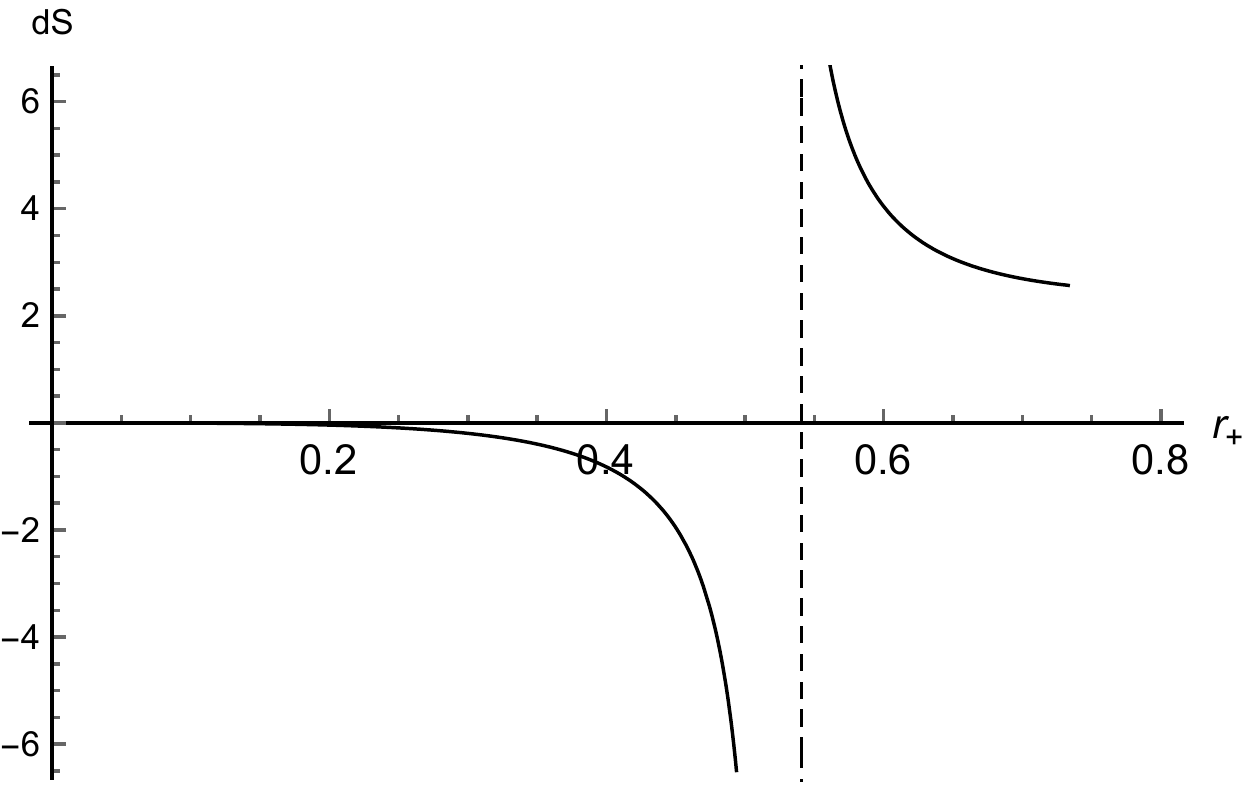}
\caption{The relation between $dS$ and $r_+$ which parameter values are $M = 0.5$, $l = 1$, $\omega_q=-2/3$, $a=0.1$, $\alpha=0.1$, $da=0.5$, $d\alpha=0.1$ and $dt=0.0001$.}
\label{fig:dS1}
\end{figure}

The relationship between $dS$ and $r_+$ can also be visualized as a function graph, which is shown in Fig. \ref{fig:dS1}. From Fig. \ref{fig:dS1}, it is obviously that there is indeed a phase change point causes a positive or negative change in the value of $dS$. When the charge of the black hole is less than the extreme value of the charge, the entropy decreases. When the charge is greater than the extreme charge, the entropy increases. Thus, the second law of thermodynamics is indefinite in the extended phase space.

\begin{figure}[htb]
\centering
\includegraphics[scale=0.8]{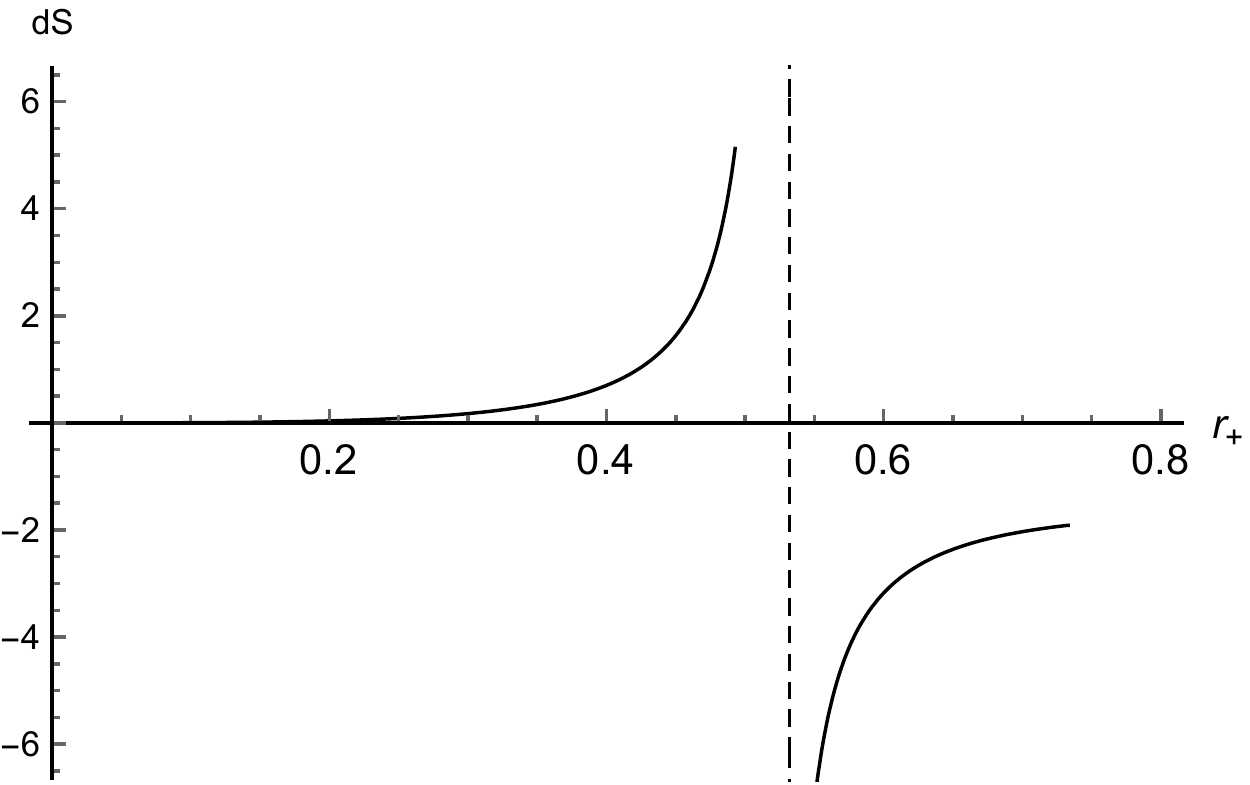}
\caption{The relation between $dS$ and $r_+$ which parameter values are $M = 0.5$, $l = 1$, $\omega_q=-2/3$, $a=0.1$, $\alpha=0.1$, $da=-0.5$, $d\alpha=0.1$ and $dt=0.0001$.}
\label{fig:dS2}
\end{figure}

In Table \ref{tab:dS2} and Fig. \ref{fig:dS2}, we set $da=-0.5$, $d\alpha=0.1$, $dt=0.0001$ and other values of variables are the same as in Table \ref{tab:dS1}. There is also a phase change point that makes the value of $dS$ positive and negative. Besides, compared to Table \ref{tab:dS1}, the sign of $dS$ corresponds to the same $r_+$ is opposite. This means that entropy increases as the charge of the black hole is less than the extreme value of the charge and decreases as the charge is greater than the extreme charge.
\begin{table}[htb]
\begin{centering}
\begin{tabular}{ccccc}
\hline
\hline
$Q $               &$r_+ $               & $dS $       &$da$      &$d\alpha$  \\
\hline
0.487372428447    & 0.420154956933     & $0.959003$      &          &              \\
0.487372          & 0.420632           & $0.966479$      &          &              \\
0.48              & 0.480899           & $3.41161$     &          &              \\
0.46              & 0.533323           & $-16.1567$      &          &              \\
0.44              & 0.565356           & $-4.91380$      &-0.5       &0.1        \\
0.40              & 0.609271           & $-2.94995$      &          &              \\
0.30              & 0.673541           & $-2.17528$      &          &             \\
0.20              & 0.708703           & $-1.99557$      &          &              \\
0.10              & 0.727240           & $-1.93008$       &          &              \\
\hline
\hline
\end{tabular}
\par\end{centering}
\caption{{\footnotesize{}{}{}{}The relation between $dS$, $Q$ and $r_+$.}}
\label{tab:dS2}
\end{table}

\begin{figure}[htb]
\centering
\includegraphics[scale=0.8]{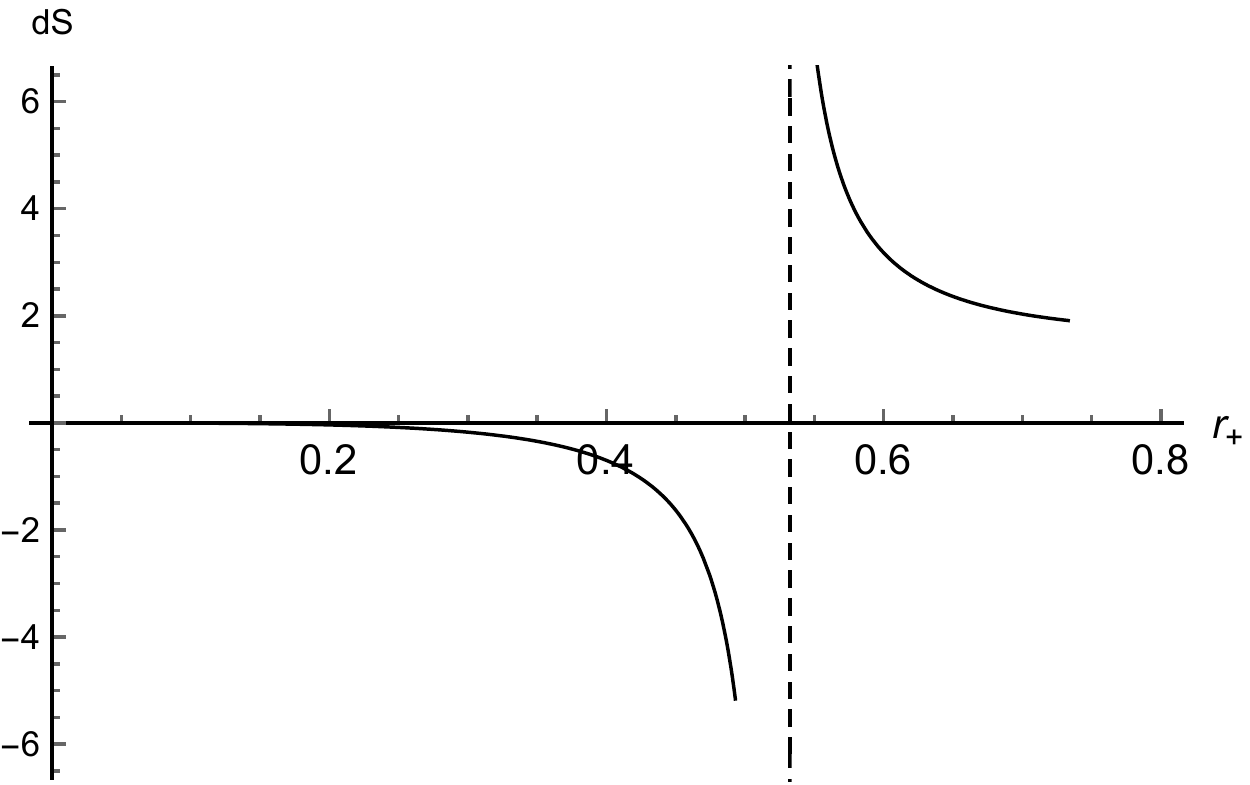}
\caption{The relation between $dS$ and $r_+$ which parameter values are $M = 0.5$, $l = 1$, $\omega_q=-2/3$, $a=0.1$, $\alpha=0.1$, $da=0.5$, $d\alpha=-0.1$ and $dt=0.0001$.}
\label{fig:dS3}
\end{figure}

In Table \ref{tab:dS3} and Fig. \ref{fig:dS3}, we set $da=0.5$, $d\alpha=-0.1$, $dt=0.0001$ and other conditions are the same as before. From Table \ref{tab:dS3}, it can obtain that for the same value of $r_+$, the absolute value of $dS$ is smaller than that in Table \ref{tab:dS1}. The remaining conclusions are the same as in Table \ref{tab:dS1} and Fig. \ref{fig:dS1}.
\begin{table}[htb]
\begin{centering}
\begin{tabular}{ccccc}
\hline
\hline
$Q $               &$r_+ $             & $dS $       &$da$      &$d\alpha$  \\
\hline
0.487372428447    & 0.420154956933     & $-0.959398$      &          &              \\
0.487372          & 0.420632           & $-0.966877$      &          &              \\
0.48              & 0.480899           & $-3.41279$     &          &              \\
0.46              & 0.533323           & $16.1612$      &          &              \\
0.44              & 0.565356           & $4.91495$      &0.5       &-0.1        \\
0.40              & 0.609271           & $2.95046$      &          &              \\
0.30              & 0.673541           & $2.17544$      &          &             \\
0.20              & 0.708703           & $1.99561$      &          &              \\
0.10              & 0.727240           & $1.93008$       &          &              \\
\hline
\hline
\end{tabular}
\par\end{centering}
\caption{{\footnotesize{}{}{}{}The relation between $dS$, $Q$ and $r_+$.}}
\label{tab:dS3}
\end{table}

From the above discussion, it can be concluded that the second law of thermodynamics is not always valid for the near-extremal black hole in the extended phase space. It is influenced by the value of $da$ and $d\alpha$. The influence of the value of $d\alpha$ is not as much as $da$.

\subsection{Weak cosmic censorship conjecture in the extended phase space}
\label{subsec:ext W}

In the extended phase space, the validity
of the weak cosmic censorship conjecture also tests through checking the sign of the minimum
value of the function $f(r)$. Assuming that there is
a minimum value of $f(r)$ at $r=r_{min}$ and the minimum value is less than zero
\begin{equation}
\delta\equiv f(r_{min})\leq0.
\end{equation}
For the extremal black hole, $\delta=0$. For the near-extremal black
hole, $\delta$ is a small quantity. After the flux of the scalar
field enters the black hole, the minimum point changes to $r_{min}+dr_{min}$.
Meanwhile, the other parameters of the black hole would change from $(M,Q,l,\alpha,a)$
to $(M+dM,Q+dQ,l+dl,\alpha+d\alpha,a+da)$. For the final state, the minimum
value of $f(r)$ is written as
\begin{equation}
\begin{aligned}
& f(r+dr,M+dM,Q+dQ,l+dl,\alpha+d\ensuremath{\alpha},a+da)|_{r=r_{min}}\\
 & =f_{min}+df_{min}\\
 & =\delta+\frac{\partial f}{\partial r}|_{r=r_{min}}dr_{min}+\frac{\partial f}{\partial M}|_{r=r_{min}}dM+\frac{\partial f}{\partial Q}|_{r=r_{min}}dQ\\
 &+\frac{\partial f}{\partial l}|_{r=r_{min}}dl+\frac{\partial f}{\partial\alpha}|_{r=r_{min}}d\ensuremath{\alpha}+\frac{\partial f}{\partial a}|_{r=r_{min}}da\\
 & =\delta+\delta_{1}+\delta_{2},
\end{aligned}
\label{eqn:ext W f1 eqn}
\end{equation}
where
\begin{equation}
\begin{aligned} & \frac{\partial f}{\partial r}|_{r=r_{min}}=0,\frac{\partial f}{\partial M}|_{r=r_{min}}=-\frac{2}{r_{min}},\frac{\partial f}{\partial Q}|_{r=r_{min}}=\frac{2Q}{r_{min}^{2}},\\
 & \frac{\partial f}{\partial l}|_{r=r_{min}}=-\frac{2r_{min}^{2}}{l^{3}},\frac{\partial f}{\partial\alpha}|_{r=r_{min}}=-\frac{1}{r_{min}^{3\omega_{q}+1}},\frac{\partial f}{\partial a}|_{r=r_{min}}=-1.\\
\end{aligned}
\label{eqn:ext W df}
\end{equation}
Inserting Eqs. $\left(\ref{eqn:ext W df}\right)$ and $\left(\ref{eqn:ext T dM}\right)$
into Eq. $\left(\ref{eqn:ext W f1 eqn}\right)$ yields
\begin{equation}
\begin{aligned} & \delta=0\\
 & \delta_{1}=-\frac{2T}{r_{min}}dS+\frac{2\left(r_{+}^{3}-r_{min}^{3}\right)}{l^{3}r_{min}}dl+\frac{r_{+}^{-3\omega_{q}}-r_{min}^{-3\omega_{q}}}{r_{min}}d\ensuremath{\alpha}+\frac{1}{r_{min}}\text{(}r_{+}-r_{min})da\\
 & \delta_{3}=-\frac{2qQr_{+}^{2}\left(\omega-q\varphi\right)}{r_{min}}\left(\frac{1}{r_{+}}-\frac{1}{r_{min}}\right)dt.
\end{aligned}
\label{eqn:ext W delta}
\end{equation}
In the extremal black hole, $r_{min}=r_{+}$, $T=0$, $f_{min}=\delta=0$ and $df_{\min}=0$. Hence, Eq. $\left(\ref{eqn:ext W f1 eqn}\right)$ is written as
\begin{equation}
f\left(M+dM,Q+dQ,l+dl,\alpha+d\alpha,a+da,r+dr\right)|_{r=r_{min}}=0.
\end{equation}
Therefore, the scattering doesn't cause changes in the minimum
value of $f(r)$. This proves that the extremal black hole is still the extremal black hole
and the weak cosmic censorship conjecture is valid. For the near-extremal black hole, $f^{\prime}(r_{+})$
is very close to zero, $f(r_{+})=0$ and $f^{\prime}(r_{min})=0$.
To calculate the value of Eq. $\left(\ref{eqn:ext W f1 eqn}\right)$,
we can suppose that $r_{+}=r_{min}+\epsilon$, where $0<\epsilon\ll1$.
In this situation, Eq. $\left(\ref{eqn:ext W delta}\right)$ is written
as
\begin{equation}
\begin{aligned}
& \delta<0\\
&\delta_{1}=-\frac{f''(r_{+})}{4\text{\ensuremath{\pi}}r_{+}}\text{\ensuremath{\epsilon}}dS+\frac{6r_{+}}{l^{3}}\text{\ensuremath{\epsilon}}dl-3\omega_{q}r_{+}^{-3\omega_{q}-2}\text{\ensuremath{\epsilon d}\ensuremath{\alpha}}-\frac{1}{r_{+}}\epsilon da+O(\epsilon^{2})\\
& \delta_{2}=\frac{2qQ(\omega-q\ensuremath{\varphi})}{r_{+}}\text{\ensuremath{\epsilon}}dt+O(\epsilon^{2}).
\end{aligned}
\end{equation}
In the above equation, $dt$ is an infinitesimal scale and
is set $dt\sim\epsilon$. For the near-extremal black hole, it is easy
to know that $dS\sim\epsilon$, $dt\sim\epsilon$, $da\sim\epsilon$
and $d\alpha\sim\epsilon$. Therefore, it is obviously that
\begin{equation}
\delta_{1}+\delta_{2}\ll\delta.
\end{equation}
Then Eq. $\left(\ref{eqn:ext W f1 eqn}\right)$ is written as
\begin{equation}
f\left(M+dM,Q+dQ,l+dl,a+da,\alpha+\text{\ensuremath{d\alpha}},r+dr\right)|_{r=r_{min}}\approx\delta<0.
\end{equation}
Therefore, the event horizon exists and the black hole isn't overcharged in the finial state. The weak cosmic censorship conjecture is valid in the near-extremal black hole.
\section{conclusion}
\label{sec:con}
In this paper, we investigated the thermodynamics and weak cosmic censorship conjecture in a RN-AdS black bole with cloud of strings and quintessence by the scattering of a scalar field. First we reviewed the thermodynamics in the black hole. Moreover, the variations of this black hole’s energy and charge within
the infinitesimal time interval are investigated. Then we calculated the
variation of the thermodynamic quantities of the black hole
after the scalar field scattering. Finally we recovered the first law of thermodynamics and discussed the validity of the second law of thermodynamics and weak cosmic censorship conjectures. We considered two cases, normal phase space and extended phase space. Our results are summarized in Table \ref{tab:12wccc}.

\begin{table}[htb]
\begin{centering}
\begin{tabular}{|p{0.9in}|p{2.7in}|p{2.6in}|}
\hline
 & Normal phase space  & Extended phase space\tabularnewline
\hline
1st law  & Satisfied  & Satisfied\tabularnewline
\hline
2nd law  & Satisfied  & Indefinite\tabularnewline
\hline
WCCC  & Satisfied for the extremal and near-extremal black holes. After the
scalar field scattering, the extremal black hole will change to a non-extremal black hole under certain conditions.  & Satisfied for the extremal and near-extremal black holes. After the
scalar field scattering, the extremal black hole stays extremal.\tabularnewline
\hline
\end{tabular}
\par\end{centering}
\caption{{\footnotesize{}{}{}{}Results for the first and second laws of
thermodynamics and the weak cosmic censorship conjectures (WCCC),
which are tested for a RN-AdS black holes with cloud of strings
and quintessence after the scalar field scattering.}}
\label{tab:12wccc}
\end{table}

In the extend phase space, there is a phase
change point that makes the the value of $dS$ positive and negative. When the value of $r_+$ is fixed, the validity of the second law of thermodynamics is determined by the value of $da$ and $d\alpha$ and the influence of $da$ is greater. In Ref. \cite{intro-He:2019fti,intro-Hong:2019yiz}, the thermodynamics and weak cosmic censorship conjecture of RN-AdS black hole only surrounded by quintessence were discussed in the normal and extended phase space.
When testing the valid of the weak cosmic censorship conjectures in the normal phase space, after the scalar field scattering, the extremal black hole still a extremal black hole when $\omega=q\varphi$ and changes to non-extremal black hole when $\omega\neq q\varphi$. In the extended phase space, the extremal black hole stays extremal after the scalar field scattering. The near-extremal black hole stays near-extremal in both phase space after the scalar field scattering.

The parameters related
to the cloud of strings and the quintessence, respectively $a$ and $\alpha$,
are considered as thermodynamic variables in this paper. Then in the extended phase space, the first law of thermodynamics
is written as
\begin{equation}
dM=TdS+VdP+\text{\ensuremath{\varphi}}dQ+\gamma\text{\ensuremath{d\alpha}}+\varkappa da,
\end{equation}
where
\begin{equation}
\gamma=-\frac{1}{2r_{+}^{3\omega_{q}}},\varkappa=-\frac{r_{+}}{2}.
\end{equation}
In previous papers, when only consider cosmological constant and
treat $a$ and $\alpha$ as constants, the first law of thermodynamics
is shown as
\begin{equation}
dM=TdS+VdP+\text{\ensuremath{\varphi}}dQ.
\end{equation}
Furthermore the expressions of $dr_{+}$, $dS$ and $f(r_{min})$
have no terms related to the parameters $a$ and $\alpha$. The influence of the cloud
of quintessence on black holes cannot be reflected under this conditions. As noted in \cite{intro-He:2019fti}, the parameters related to the quintessence is treated as constants and there is no terms related to the parameter related to the quintessence. Therefore, to study the effects
of cloud of strings and quintessence on black holes, we need to consider the parameters
$a$ and $\alpha$ during the discussion of thermodynamics and weak cosmic censorship conjecture. It is confirmed that the cloud of strings and quintessence both have certain effects on the thermodynamic quantities of black holes. The results of the first
thermodynamic law under different conditions are summarized in Table \ref{tab:1st}.
\begin{table}[htb]
\begin{centering}
\begin{tabular}{|p{3.0in}|p{3.0in}|}
\hline
Types of black holes & 1st law  \tabularnewline
\hline
RN-AdS BH  & $dM=TdS+VdP+\text{\ensuremath{\varphi}}dQ$\tabularnewline
\hline
RN-AdS BH with cloud of strings  & $dM=TdS+VdP+\text{\ensuremath{\varphi}}dQ-\frac{r_{+}}{2}da$\tabularnewline
\hline
RN-AdS BH with quintessence  & $dM=TdS+VdP+\text{\ensuremath{\varphi}}dQ-\frac{1}{2r_{+}^{3\omega_{q}}}d\alpha$\tabularnewline
\hline
RN-AdS BH with cloud of strings and quintessence  & $dM=TdS+VdP+\text{\ensuremath{\varphi}}dQ-\frac{1}{2r_{+}^{3}\omega_{q}}d\alpha-\frac{r_{+}}{2}da$\tabularnewline
\hline
\end{tabular}
\par\end{centering}
\caption{{\footnotesize{}{}{}{}Results for the first
thermodynamic law under different conditions.}}
\label{tab:1st}
\end{table}

\end{document}